\documentclass{svproc}
\usepackage[numbers,sectionbib]{natbib}
\usepackage{url}

\usepackage{amsmath}
\usepackage{algorithm}
\usepackage{algorithmicx}
\usepackage{amsmath,algorithm,tabularx}
\usepackage{algpseudocode}
\usepackage{amssymb}
\usepackage{multirow}
\usepackage{makecell}
\usepackage{caption}
\usepackage{array}
\usepackage[pdftex]{graphicx}

\newcolumntype{P}[1]{>{\centering\arraybackslash}p{#1}}
\newcolumntype{C}{>{\centering\arraybackslash} m{6cm} }
\makeatletter
\newcommand{\multiline}[1]{%
	\begin{tabularx}{\dimexpr\linewidth-\ALG@thistlm}[t]{@{}X@{}}
		#1
	\end{tabularx}
}
\makeatother

\begin{document}
\mainmatter              % start of a contribution
\title{Deep-Reinforcement-Learning-Based Scheduling with Contiguous Resource Allocation for Next-Generation Wireless Systems}
\titlerunning{Deep-Reinforcement-Learning-Based Scheduling}  % abbreviated title (for running head)

\author{Shu Sun \and Xiaofeng Li}
\authorrunning{Sun et al.} % abbreviated author list (for running head)
\tocauthor{Ivar Ekeland, Roger Temam, Jeffrey Dean, David Grove,
	Craig Chambers, Kim B. Bruce, and Elisa Bertino}
\institute{Intel Corporation, Santa Clara, CA 95054, USA,\\
	\email{shu.sun@intel.com, xiaofeng.li@intel.com}}

\maketitle              % typeset the title of the contribution

\begin{abstract}
Scheduling plays a pivotal role in multi-user wireless communications, since the quality of service of various users largely depends upon the allocated radio resources. In this paper, we propose a novel scheduling algorithm with contiguous frequency-domain resource allocation (FDRA) based on deep reinforcement learning (DRL) that jointly selects users and allocates resource blocks (RBs). The scheduling problem is modeled as a Markov decision process, and a DRL agent determines which user and how many consecutive RBs for that user should be scheduled at each RB allocation step. The state space, action space, and reward function are delicately designed to train the DRL network. More specifically, the originally quasi-continuous action space, which is inherent to contiguous FDRA, is refined into a finite and discrete action space to obtain a trade-off between the inference latency and system performance. Simulation results show that the proposed DRL-based scheduling algorithm outperforms other representative baseline schemes while having lower online computational complexity. 
\end{abstract}
\section{Introduction}
Resource allocation is an indispensable ingredient in wireless systems with multiple user equipments (UEs), in order to meet certain quality of service (QoS) requirements such as throughput, fairness, latency, and/or reliability. The 3rd Generation Partnership Project (3GPP) has specified two types of downlink frequency-domain resource allocation (FDRA), i.e., type 0 and type 1, for the fifth-generation (5G) and beyond-5G (B5G) wireless communications~\cite{38214,38331}. There exist two essential discrepancies between type-0 and type-1 FDRA: (1) Type 0 is on the resource block group (RBG) level, where an RBG contains a number of consecutive resource blocks (RBs) and an RB is defined as 12 consecutive subcarriers in the frequency domain~\cite{38211}, while type 1 is on the RB level; (2) The resources (RBGs or RBs) assigned to each UE can be non-contiguous for type 0, while they must be contiguous for type 1. 

The contiguous-RB constraint in type-1 FDRA renders it extremely difficult to find an optimal UE and resource allocation strategy except using brute-force search whose computational complexity is prohibitively high. Consequently, it is vital to propose sub-optimal scheduling algorithms with contiguous FDRA that have reasonable complexity hence implementable in practice. A myriad of contiguous FDRA scheduling approaches have been proposed previously (for instance,~\cite{Tsiropoulou16,Wong11,Sun20Fdra} and references therein). In particular, three practical scheduling algorithms with contiguous FDRA have been presented in~\cite{Sun20Fdra}, wherein one of the algorithms, joint allocation with dual ends (JADE), yields the best performance and outperforms an existing contiguous FDRA method representative in the industry~\cite{Wong11,Sun20Fdra}. Nevertheless, albeit its relatively low complexity compared with prior schemes, JADE still incurs considerable complexity when UEs abound since its complexity scales with the square of the number of UEs. On the other hand, artificial intelligence (AI) has found a variety of applications in the field of wireless communications~\cite{Li18Asilomar,Luong19,Li19Asilomar,Yan20Access,Zhang19CST,Yan20IoT}, which can help solve problems difficult to handle using conventional non-AI methods, improve efficiency and/or performance of existing solutions, or reduce instantaneous computation time while offering comparable performance, among other advantages. Therefore, it will be beneficial if AI techniques can be leveraged to schedule UEs and resources that yields performance comparable to or even better than JADE while inducing noticeably lower online computational complexity. 

Given the fact that UE and resource scheduling can be modeled by a Markov decision process~\cite{Luong19,Puterman14}, deep reinforcement learning (DRL), an important branch of machine learning belonging to AI, can be adopted to train an agent to offer optimal or superhuman performance while requiring significantly less instantaneous computational effort. DRL has been applied to solve resource allocation problems in a wide range of fields including Internet of Things~\cite{He20TETC}, vehicular communications~\cite{Ye19TVT}, heterogeneous cellular networks~\cite{Zhao19TWC}, and cloud radio access network~\cite{Xu17ICC}, to name a few. None of the existing works, however, has tackled FDRA with the contiguity constraint which is a crucial resource allocation approach in the 3GPP specifications for 5G and B5G systems~\cite{38214,38331,38211}.

In this work, we propose a novel algorithm that jointly schedules UEs and contiguous frequency-domain resources based on DRL, which is named as STAR (Super Type-1 Allocation based on Reinforcement learning). Specifically, after trained offline, the DRL-based algorithm can make judicious decisions instantaneously on which UE and how many RBs for that UE shall be scheduled jointly at each allocation step. Major merits of the proposed algorithm are two-fold: (1) it can yield remarkable performance which is even superior to JADE, and (2) it enjoys considerably reduced online computational complexity as compared to JADE. To the authors' best knowledge, this is the first work that utilizes DRL to perform scheduling with contiguous FDRA. 

\section{System Model and Problem Formulation}
We investigate a downlink cellular system comprising one next-generation nodeB (gNB) and $K$ UEs indexed by the set $\mathcal{K}=\{0,...,K-1\}$, where the UEs' traffic types can be diverse with distinct QoS requirements. The transmission bandwidth part (BWP) $W$ is orthogonally divided into $B$ RBs indexed by the set $\mathcal{B}=\{0,...,B-1\}$. The payload for UE $k$ is denoted by $L_k$. Two constraints exist in the aforementioned type-1 FDRA in 3GPP 5G and B5G specifications~\cite{38214,38331}: (1) exclusivity, i.e., an RB can only be allocated to at most one UE in the same time resource; (2) contiguity, indicating that the RBs assigned to each UE must be contiguous.  

For UE $k$ on RB $b$, given the estimated channel $\mathbf{H}_{k,b}$ and precoding matrix codebook~\cite{38214}, the rank indicator (RI), precoding matrix indicator (PMI), and modulation and coding scheme (MCS)~\cite{38214} can be obtained, e.g., using the scheme in~\cite{Sun20}, after which the transport block size (TBS) per slot, $\mathrm{TBS}_{k,b}$, is computed based on the RI, PMI, and MCS. The achievable rate of UE $k$ on RB $b$ in each slot is $r_{k,b}=\mathrm{TBS}_{k,b}$. Let $\mathcal{B}_k$ denote the set of RBs allocated to UE $k$, the achievable rate of UE $k$ over $\mathcal{B}_k$ is $r_k=\sum_{b \in\mathcal{B}_k}r_{k,b}$. The scheduling metric (e.g., sum-rate, proportional fairness (PF)~\cite{Tse99}, among others) can be flexible depending upon the system requirement. In this paper, sum-rate is selected as the scheduling metric as an example. The optimization problem can be formulated as  
\begin{equation}\label{eq1}
	\begin{alignedat}{2}
		\text{(P1):}~ &\max_{\{\mathcal{B}_0,...,\mathcal{B}_{K-1}\}\subseteq\mathcal{D}}&&\sum_{k \in\mathcal{K}}\sum_{b \in\mathcal{B}_k}r_{k,b} \\
		&\text{subject~to} & & \mathcal{B}_k\cap\mathcal{B}_{k^\prime}=\emptyset,\forall k\ne k^\prime, k,k^\prime\in\mathcal{K}, \\
		& & &d_k\leq\tau_k, \forall k\in\mathcal{K}
	\end{alignedat}
\end{equation}

\noindent where $\mathcal{D}$ is the set of all possible RB allocations satisfying the contiguity constraint, $d_k$ and $\tau_k$ denote the head-of-line (HoL) delay and delay threshold (the maximum allowable delay from packet generation to packet scheduling) of UE $k$, respectively. Note that a packet will be dropped and will not contribute to the TBS if it is not entirely scheduled before its HoL delay exceeds its delay threshold. The optimal solution to (P1) requires exhaustive search with prohibitively high computational complexity. In~\cite{Sun20Fdra}, a sub-optimal algorithm JADE has been put forth to solve (P1), which has been proved to provide near-optimal performance based on simulation results. The procedures of JADE is detailed in Algorithm 1. The main design principle of JADE is to jointly prioritize the UE and RB(s) in each allocation step that produces the largest scheduling metric with the minimum number of RBs, where the RB selection is performed and compared between both ends of the active BWP to exploit frequency diversity. More specifically, for each UE, JADE first calculates the numbers of RBs needed to transmit its payload from both ends of the BWP, and selects the end of the BWP that requires fewer RBs, and computes the associated scheduling metric (i.e., sum-TBS herein). Then the UE possessing the largest scheduling metric is selected and its final MCS is calculated over the selected RBs. The steps above are executed iteratively until there is no remaining UE or RB. 
\begin{algorithm}
	\caption{Joint Allocation with Dual Ends (JADE)}\label{algo:JADE}
	\begin{algorithmic}[1]% The number tells where the line numbering should start
		\Require {Initialize $\mathcal{K}^\star=\emptyset$, $\mathcal{B}^\star=\emptyset$.}
		\While{$\mathcal{K}\neq\emptyset$ and $\mathcal{B}\neq\emptyset$} %\Comment{We have the answer if r is 0}
		\For{$\forall k\in\mathcal{K}$}
		\State \multiline{%
			Calculate the number of RBs needed, $n_{k,\text{start}}$, to transmit $L_k$ starting from the first remaining RB in $\mathcal{B}$ and going forward, until $r_{k,\text{start}}\geq L_k$ or $\mathcal{B}=\emptyset$. Denote the selected RB set as $\mathcal{B}_{k,\text{start}}$.}
		\State \multiline{%
			Calculate the number of RBs needed, $n_{k,\text{end}}$, to transmit $L_k$ starting from the last remaining RB in $\mathcal{B}$ and going backward, until $r_{k,\text{end}}\geq L_k$ or $\mathcal{B}=\emptyset$. Denote the selected RB set as $\mathcal{B}_{k,\text{end}}$.}
		\State \multiline{%
			If $n_{k,\text{start}}\leq n_{k,\text{end}}$, store $\mathcal{B}_{k,\text{start}}$ and $r_{k,\text{start}}$ as $\mathcal{B}_k$ and $r_k$, respectively; otherwise store $\mathcal{B}_{k,\text{end}}$ and $r_{k,\text{end}}$ as $\mathcal{B}_k$ and $r_k$, respectively.}
		\EndFor
		\State \multiline{%
			$k^\star=\underset{k}{\mathrm{argmax}}~r_k$.}
		\State \multiline{%
			Calculate $\text{MCS}_{k^\star}$, the final MCS for UE $k^\star$ over $\mathcal{B}_{k^\star}$.}
		\State \multiline{%
			$\mathcal{K}^\star\gets\mathcal{K}^\star\cup\{k^\star\}$, $\mathcal{B}^\star\gets\mathcal{B}^\star\cup\mathcal{B}_{k^\star}$.}
		\State \multiline{%
			$\mathcal{K}\gets\mathcal{K}\setminus\{k^\star\}$, $\mathcal{B}\gets\mathcal{B}\setminus\mathcal{B}_{k^\star}$.}
		\EndWhile
		%\State \textbf{return} $\mathcal{K}^\star$ and $\mathcal{B}^\star$.  
		\State \textbf{return} $\mathcal{K}^\star$, $\mathcal{B}^\star$, and $\text{MCS}_k,\forall k\in\mathcal{K}^\star$. 
	\end{algorithmic}
\end{algorithm}
\begin{algorithm}
	\caption{Super Type-1 Allocation based on Reinforcement learning (STAR)}\label{algo:STAR}
	\begin{algorithmic}[1]% The number tells where the line numbering should start
		\Require {Initialize $\mathcal{K}^\star=\emptyset$, $\mathcal{B}^\star=\emptyset$, $\textbf{s}$ (to be detailed in Algorithm~\ref{algo:starState}).}
		\While{$\mathcal{K}\neq\emptyset$ and $\mathcal{B}\neq\emptyset$} %\Comment{We have the answer if r is 0}
		\State \multiline{%
			The DRL agent takes an action $a$ based upon $\textbf{s}$, where $a$ indicates which UE and how many RBs should be allocated (see Table~\ref{tbl:action} for more detailed information). Denote the selected UE and selected RB set as $k^\star$ and $\mathcal{B}_{k^\star}$, respectively.}
		\State \multiline{%
			The DRL agent passes $a$ to the environment.}
		\State \multiline{%
			The environment generates a reward $r$ and the next state $\tilde{\textbf{s}}$ based on $a$, and passes them to the DRL agent.}
		\State \multiline{%
			Calculate $\text{MCS}_{k^\star}$, the final MCS for UE $k^\star$ over $\mathcal{B}_{k^\star}$.}
		\State \multiline{%
			$\textbf{s}\gets\tilde{\textbf{s}}$.}
		\State \multiline{%
			$\mathcal{K}^\star\gets\mathcal{K}^\star\cup\{k^\star\}$, $\mathcal{B}^\star\gets\mathcal{B}^\star\cup\mathcal{B}_{k^\star}$.}
		\State \multiline{%
			$\mathcal{K}\gets\mathcal{K}\setminus\{k^\star\}$, $\mathcal{B}\gets\mathcal{B}\setminus\mathcal{B}_{k^\star}$.}
		\EndWhile
		%\State \textbf{return} $\mathcal{K}^\star$ and $\mathcal{B}^\star$.  
		\State \textbf{return} $\mathcal{K}^\star$, $\mathcal{B}^\star$, and $\text{MCS}_k,\forall k\in\mathcal{K}^\star$. 
	\end{algorithmic}
\end{algorithm}

\section{Proposed Scheduling Algorithm Based on DRL}
\begin{figure}
	\centering
	\includegraphics[width=\columnwidth]{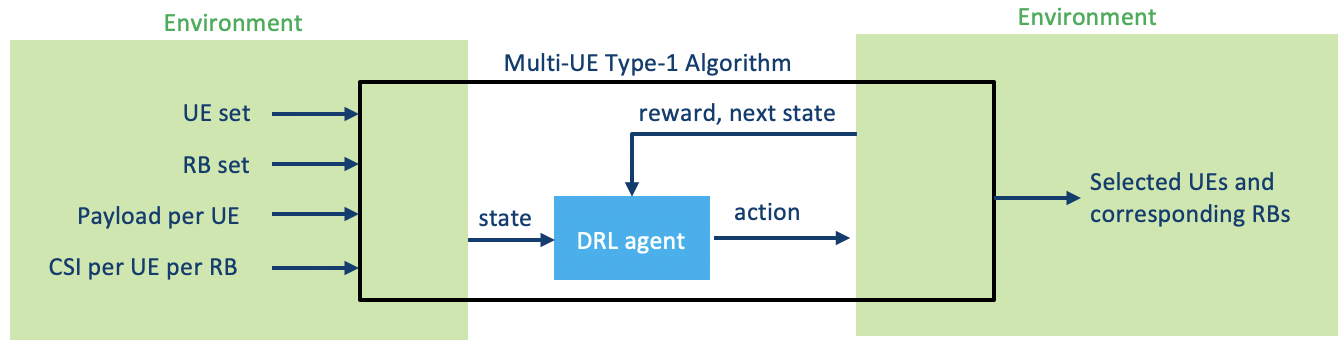}
	\caption{Input and output relationship per slot for multi-UE type-1 FDRA based on DRL. The green shaded modules represent the environment, and the blue module denotes the DRL agent.}
	\label{fig:inputOutput}	
\end{figure}

It is noteworthy that in JADE, the number of scheduling metric calculation is proportional to $K^2$ due to the iteration for each remaining UE and RB. To reduce the online computational complexity and further enhance performance, we propose using a DRL-based scheduling approach, STAR, to smartly allocate RBs to multiple UEs. A deep Q-network (DQN) is employed for the training of STAR. The input and output relationship per slot incorporating DRL is illustrated in Fig.~\ref{fig:inputOutput}, where the green shaded modules represent the environment, and the blue module depicts the DRL agent. Each slot usually includes multiple allocation steps, wherein at each allocation step, a UE and a number of consecutive RBs are selected to be scheduled. Within each slot, a state is generated from the environment and is then passed to the DRL agent, after which the DRL agent takes an action and outputs to the environment, then a reward and the next state are produced from the environment based on the current state and action pair, and the steps above are repeated multiple times until there is no remaining UE or RB. The overall process of STAR is detailed in Algorithm~\ref{algo:STAR}. Theoretically, the RB allocation involves the determination of both the starting location and number of RBs, which is a highly entangled problem. To reduce the complexity of the DRL training and inference without significantly compromising the TBS performance, the starting location of RBs to be allocated is fixed as the first remaining RB in the BWP, so that each action only needs to determine which UE and how many RBs should be allocated without dealing with the RB starting location. The inclusion of the starting location of RB allocation is deferred to future work. 

\subsection{State Design}
In this work, as mentioned previously, DRL is utilized to determine which UE and how many RBs for the selected UE should be scheduled at each allocation step. The state design is closely related to the scheduling metric used in the scheduling algorithm. Since the scheduling metric adopted herein is sum-rate, or equivalently, sum-TBS, both RB-related channel state information (CSI) and the metric related to sum-TBS (i.e., the remaining payload), should be reflected in the state. Detailed methodology for the acquisition of the current state $\textbf{s}$ at each RB allocation step is given in Algorithm~\ref{algo:starState}. It is worth noting that if the next UE scheduling and resource allocation takes place in the current slot, the next state $\tilde{\textbf{s}}$ is the state after the current UE scheduling and resource allocation in the current slot. In contrast, if the next UE scheduling and resource allocation occurs in the next slot, $\tilde{\textbf{s}}$ takes the first state in the next slot, which reflects both the decision made in the current slot and new packet information and CSI in the next slot. This way, the state transition is always continuous and Markovian for both intra-slot and inter-slot cases. 
\begin{algorithm}
	\caption{State Design for STAR}\label{algo:starState}
	\begin{algorithmic}[1]% The number tells where the line numbering should start
		\For{$\forall k\in\mathcal{K}$}
		\For{$\forall b\in\mathcal{B}$}
		\State \multiline{%
			$g_{k,b}=\text{MCS}_{k,b}\times\text{RI}_{k,b}$}
		\EndFor
		\If{$L_k>0$}
		\State \multiline{%
			$\textbf{s}_k=[L_k,g_{k,0},g_{k,1},...,g_{k,B-1}]$}
		\Else
		\State \multiline{%
			$\textbf{s}_k=\underbrace{[-1,-1,-1,...,-1]}_{B+1}\}$}
		\EndIf
		\EndFor
		\State \multiline{%
			$\textbf{s}=[\textbf{s}_0,\textbf{s}_1,...,\textbf{s}_{K-1}]$}
		\State \textbf{return} $\textbf{s}$
	\end{algorithmic}
\end{algorithm}

\subsection{Action Design}
For contiguous FDRA, the action consists of two aspects: UE selection and RB allocation, which leads to a quasi-continuous action space $\mathcal{A}$ due to potentially large numbers of RBs to be allocated. Considering the trade-off between training/inference overhead and performance, we transform the quasi-continuous action space into a discrete and finite action space. In particular, the size of the action space herein is devised to be $|\mathcal{A}|=5K$, where the actions are indexed by $0,1,...,5K-1$. The quotient produced by the division of the action index over 5 equals the UE index, i.e., $a//5=k$, where $a$ denotes the action index. For example, Actions 0 to 4 refer to selecting UE $0$, Actions 5 and 9 indicate selecting UE $1$, so on and so forth. In order to determine how many RBs should be allocated to each UE, the number of RBs needed to transmit the payload $L_k$, denoted as $n_{k,\text{WB}}$, is first calculated based on the wideband (WB) channel quality indicator (CQI)~\cite{38214} of each UE, where the procedure in Section II of~\cite{Sun20Fdra} is adopted to obtain the WB CQI. The number of RBs allocated to UE $k$, $n_k$, is given by 
\begin{equation}\label{eq:action}
	n_k=n_{k,\text{WB}}+a\%5-2
\end{equation}

\noindent where $a\%5$ represents the remaining resultant from the division of $a$ over 5. Intuitively, the five actions associated with UE $k$ are related to the number of RBs to be allocated based on $n_{k,\text{WB}}$. The overall actions and their meanings are listed in Table~\ref{tbl:action}.
\begin{table}[!t]
	\renewcommand{\arraystretch}{1.2}
	\caption{Action $a$ and the Corresponding Physical Meaning}
	\label{tbl:action}
	\centering
	\begin{tabular}{|c||c|c|c|c|c|}
		\hline
		$a//5$ & \multicolumn{5}{c|}{$k$} \\
		\hline
		\makecell{Selected UE} & \multicolumn{5}{c|}{$k$} \\
		\hline
		$a\%5$ & 0 & 1 & 2 & 3 & 4 \\
		\hline
		\makecell{$n_k$}& \makecell{$n_{k,\text{WB}}-2$} & \makecell{$n_{k,\text{WB}}-1$} & $n_{k,\text{WB}}$ & \makecell{$n_{k,\text{WB}}+1$} & \makecell{$n_{k,\text{WB}}+2$}\\
		\hline
	\end{tabular}
\end{table}

Note that variants of the action design described above can also be considered in practice according to specific overhead and/or performance requirements. For instance, the range of actions per UE can be smaller or larger than $5K$, i.e., $n_k$ can be within $\pm1$ with respect to $n_{k,\text{WB}}$ to accelerate training and inference processes, or $\pm3,\pm4,...$ with respect to $n_{k,\text{WB}}$ to improve scheduling performance. Furthermore, the range of actions per UE can be asymmetric with respect to $n_{k,\text{WB}}$. 

\subsection{Reward Design}
As the scheduling metric is sum-TBS per slot, the reward should incarnate the allocated total TBS in a slot. Therefore, at each allocation step $i$, the allocated TBS per UE is first calculated, denoted as $\text{TBS}_{k,i}$ for UE $k$. Next, a temporary quantity $p_i$ for allocation step $i$ is computed as
\begin{equation}\label{eq:p}
	p_i=\text{min}\left(\sum_{k=1}^{K}\text{TBS}_{k,i},\sum_{k=1}^{K}L_k\right)\Bigg/\sum_{k=1}^{K}L_k
\end{equation}

\noindent whose physical meaning is the normalized transmitted sum-rate. The ultimate reward at allocation step $i$, $r_i$, is the summation of $p_j$'s associated with all the allocation steps up to allocation step $i$, i.e.
\begin{equation}\label{eq:reward}
	r_i=\sum_{j=1}^{i}p_j
\end{equation}

\noindent Since $\sum_{j=1}^{i}\sum_{k=1}^{K}\text{TBS}_{k,j}\leq\sum_{k=1}^{K}L_k$, the reward $r_i$ always falls between 0 and 1, which facilitates the training of the DQN. 

\subsection{Complexity Analysis}
The most prominent advantages of the proposed STAR algorithm over JADE are its superior performance (to be shown later in this work) and its significantly reduced online computational complexity. Assuming each UE needs $M$ RBs on average to send its payload, then the total computational complexity of JADE is around $MK^2$~\cite{Sun20Fdra}, since it necessitates the calculations of the TBS from both ends of the BWP for each UE, and repetitions of the above procedure after each UE and resource allocation stage. In contrast, by using STAR, the computational complexity at each allocation step is only 1, since it directly determines which UE and how many RBs should be scheduled with only one-time TBS calculation, hence substantially reducing the complexity. Therefore, STAR can provide about $MK^2$ times, which can be up to four orders of magnitude, of complexity reduction as compared to JADE. A brief analysis and comparison of the online computational complexity of the two algorithms is provided in Table~\ref{tbl:complexity}. 
\begin{table}[!t]
	\renewcommand{\arraystretch}{1.1}
	\caption{Comparison of Online Computational Complexity}
	\label{tbl:complexity}
	\centering
	\begin{tabular}{|c||c|c|}
		%{|p{1.9cm}||p{0.75cm}|p{0.75cm}|p{0.75cm}|p{0.75cm}|p{1.1cm}|}
		\hline
		Algorithm & JADE & STAR \\
		\hline
		\makecell{Number of scheduling metric calculation} & $MK^2$ & 1 \\
		\hline
		\makecell{Sum complexity} & $MK^2$ & 1\\
		\hline
		\makecell{Sum complexity for the case of\\$K=30$, $B=270$, $M=12$, $M_\text{RB}=4$} & \makecell{$\mathcal{O}(1e4)$} & 1\\
		\hline
	\end{tabular}
\end{table}

\subsection{Training of DQN}
In the DRL framework of this work, the agent is a DQN model composed of three fully-connected hidden layers where the number of neurons per layer is 1024, 256, and 128, respectively, in addition to an input layer and an output layer. The input size is $K(B+1)$ which is the state dimension, and the output size is the cardinality of the action space, i.e. 5$K$. The environment incorporates the wireless channel models as well as transmission and reception procedures based on the 3GPP 5G new radio (NR) standards such as~\cite{38214,38331,38211,38213,38901}, among others. In each training step, the following processes take place (as shown by Fig.~\ref{fig:inputOutput}): (1) The environment generates the current state $\textbf{s}$ and passes it to the DRL agent; (2) The DRL agent generates an action $a$ and feeds it back to the environment; (3) The environment yields the reward $r$ and the next state $\tilde{\textbf{s}}$ based on $a$, and outputs them to the DRL agent; (4) The sequence of state, action, reward, and next state, $[\textbf{s},a,r,\tilde{\textbf{s}}]$, is used to train the DQN, i.e., the agent. Experience replay and $\epsilon$-greedy action selection are performed to facilitate the training~\cite{Luong19}.

\section{Numerical Results}
System-level simulations are carried out to evaluate the performance of the proposed algorithm. Table~\ref{tbl:simSet} lists the simulation settings, where the traffic models are remoteDrivingDl (RDD) and powerDist2 (PD2) which represent the downlink remote driving scenario, and the second type of power distribution grid fault and outage management~\cite{38824}, respectively, both of which belong to URLLC (Ultra-Reliable Low-Latency Communications), one of the three major 5G usage scenarios~\cite{38824}. The total number of UEs in our simulations is $K=5$, thus yielding 25 actions. Table~\ref{tbl:traffic} details the parameters for the traffic models used in our simulations. The training and inference processes of the proposed DRL is implement in PyCharm. The relevant simulation parameter values of the DRL are given in Table~\ref{tbl:drlSimSet}.
\begin{table}[!t]
	\renewcommand{\arraystretch}{1.2}
	\caption{Simulation settings}
	\label{tbl:simSet}
	\centering
	\begin{tabular}{|c||c|}
		\hline
		Configuration & Value \\
		\hline \hline
		Transmit power & 23 dBm \\
		\hline
		Number of gNB antennas & 4 \\
		\hline
		Cell radius & 250 m \\
		\hline
		UE distribution & Uniform \\
		\hline
		Number of antennas per UE & 4 \\
		\hline
		Number of UEs per gNB & 5 \\
		\hline
		Channel& \makecell{EPA (Extended Pedestrian A model)} \\
		\hline
		Numerology& \makecell{30kHz sub-carrier spacing, 100MHz bandwidth} \\
		\hline
		CSI feedback delay & 1 slot \\
		\hline
		Traffic model& \makecell{remoteDrivingDl (RDD), powerDist2 (PD2)} \\
		\hline
	\end{tabular}
\end{table}
\begin{table}[!t]
	\renewcommand{\arraystretch}{1.2}
	\caption{Parameters for the traffic models used in simulations~\cite{38824}}
	\label{tbl:traffic}
	\centering
	\begin{tabular}{|c||c|c|}
		\hline
		& remoteDrivingDl (RDD) & powerDist2 (PD2)\\
		\hline \hline
		\makecell{Delay threshold (ms)}& 1 & 1 \\
		\hline
		\makecell{Acceptable packet drop probability}& 0.0001\% & 0.0001\%\\
		\hline
		\makecell{Packet size (bits)}& 16664 & 2000\\
		\hline
	\end{tabular}
\end{table}
\begin{table}[!t]
	\renewcommand{\arraystretch}{1.2}
	\caption{Simulation parameters of the DRL}
	\label{tbl:drlSimSet}
	\centering
	\begin{tabular}{|c||c|}
		\hline
		Parameter & Value \\
		\hline \hline
		Number of states & 255 \\
		\hline
		Number of actions & 25 \\
		\hline
		Number of hidden layers & 3 \\
		\hline
		Number of neurons per hidden layer & 1024, 256, 128 \\
		\hline
		Initial weight value & Normal initialization \\
		\hline
		Optimization algorithm & Adam \\
		\hline
		Activation function & ReLU \\
		\hline
		Epsilon decay rate & 0.996 \\
		\hline
		Learning rate & 1e-6 \\
		\hline
		Experience memory size & 16400$\times$536 \\
		\hline
		Batch size & 1024 \\
		\hline
	\end{tabular}
\end{table}
\begin{figure*}
	\centering
	\includegraphics[width=\columnwidth]{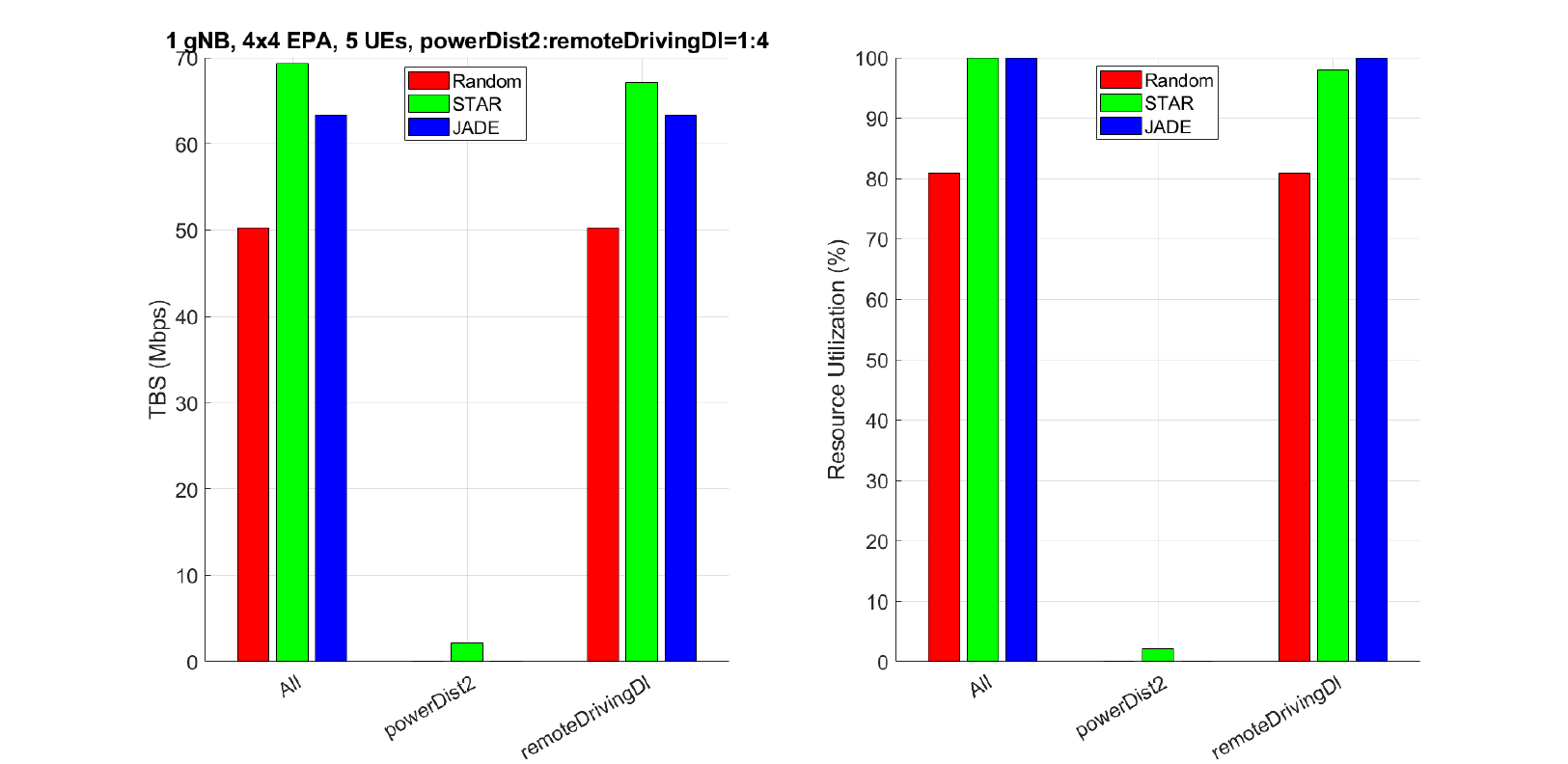}
	\caption{Performance of JADE, STAR, and random scheduling approaches. The ratio of powerDist2 and remoteDrivingDl UEs is 1:4.}
	\label{fig:pfm1}	
\end{figure*}
\begin{figure*}
	\centering
	\includegraphics[width=\columnwidth]{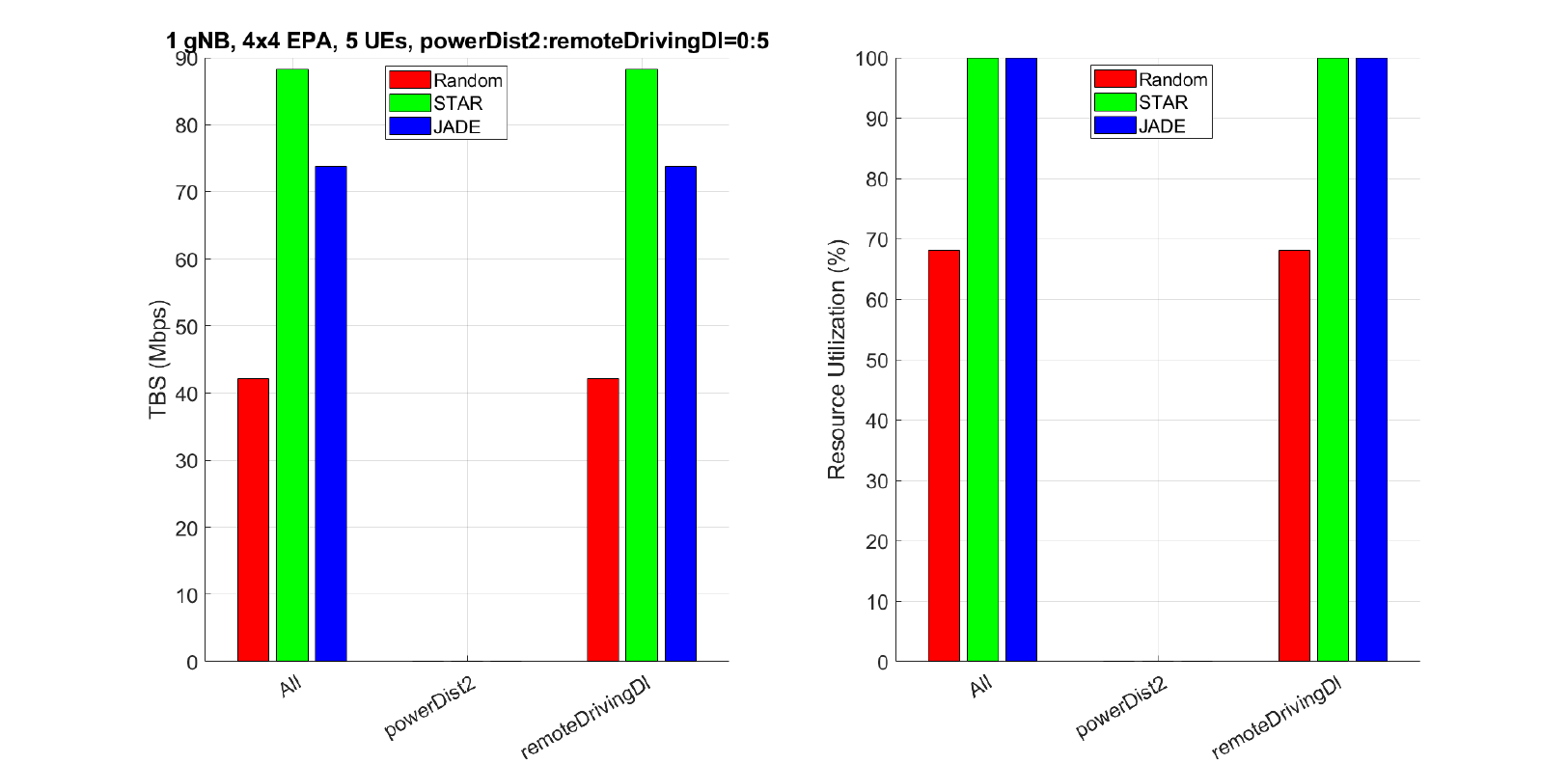}
	\caption{Performance of JADE, STAR, and random scheduling approaches. The ratio of powerDist2 and remoteDrivingDl UEs is 0:5, and the RL model used for the simulation is with a ratio of powerDist2 and remoteDrivingDl UEs of 1:4.}
	\label{fig:pfm2}	
\end{figure*}
\begin{figure*}
	\centering
	\includegraphics[width=\columnwidth]{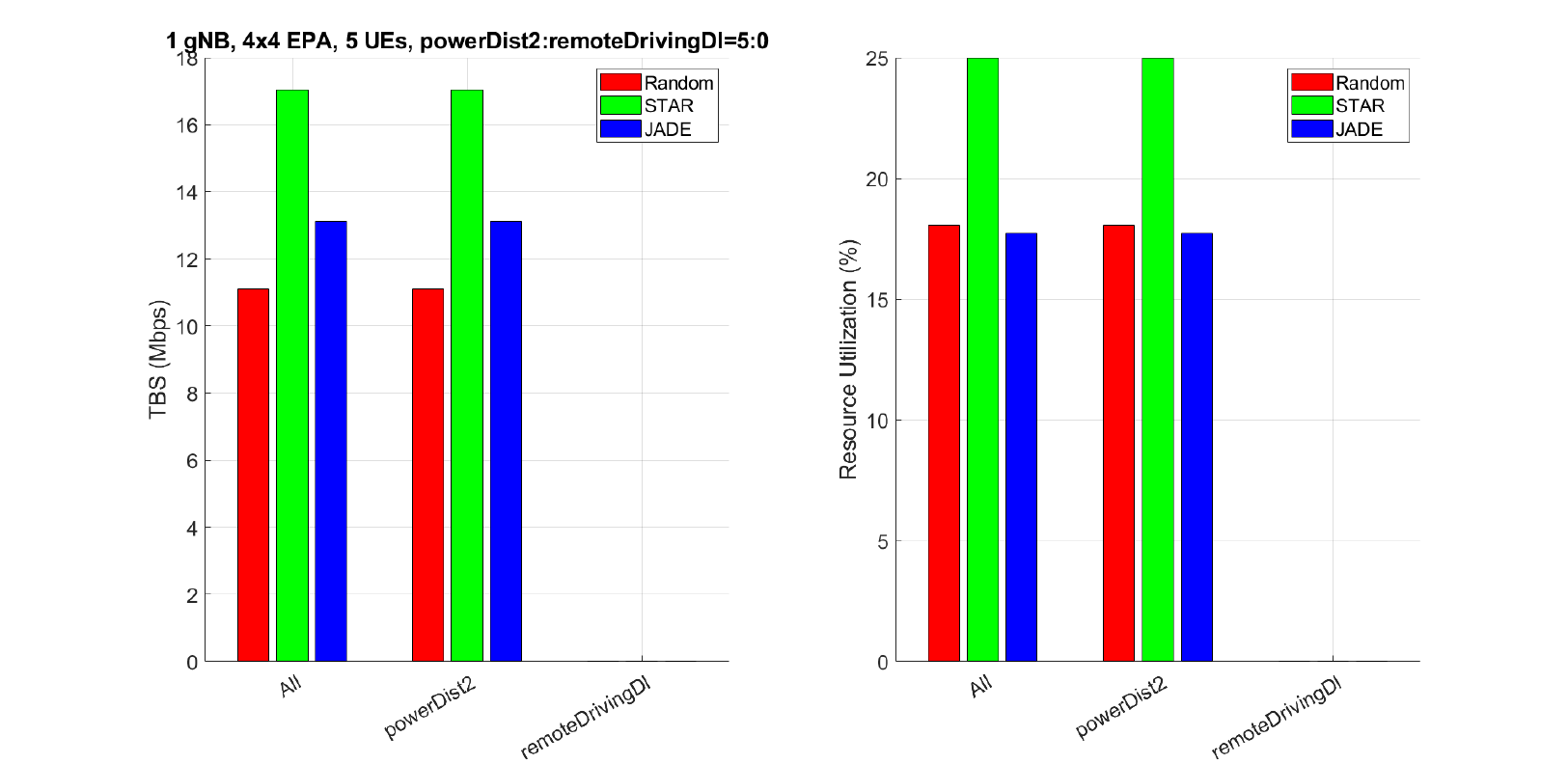}
	\caption{Performance of JADE, STAR, and random scheduling approaches. The ratio of powerDist2 and remoteDrivingDl UEs is 5:0, and the RL model used for the simulation is with a ratio of powerDist2 and remoteDrivingDl UEs of 1:4.}
	\label{fig:pfm3}	
\end{figure*}

The performance of STAR is compared against JADE and a random scheduling strategy, i.e., arbitrarily selecting which UE and how many RBs to be scheduled at each allocation step. The sum-TBS and resource utilization of these three algorithms are illustrated in Fig.~\ref{fig:pfm1}, where the resource utilization is defined as the ratio of the number of consumed RBs to the total number of RBs, and the ratio of PD2 and RDD UEs is 1:4. Moreover, Fig.~\ref{fig:pfm2} and Fig.~\ref{fig:pfm3} show the performance with different UE ratios, using exactly the same offline-trained DRL model for the case in which the ratio of PD2 and RDD UEs is 1:4. The following key observations can be drawn from Figs.~\ref{fig:pfm1}-\ref{fig:pfm3}:

\noindent 1) The TBS comparison in Fig.~\ref{fig:pfm1} unveils that STAR outperforms both JADE and random scheduling for both overall mixed and individual traffic types, demonstrating its superiority over the two schemes. Specifically, the TBS gain of STAR over JADE is about 10\% for overall mixed traffic types, 6\% for the RDD traffic, and virtually infinity for the PD2 traffic since JADE does not yield any TBS for PD2. Furthermore, the resource utilization comparison in Fig.~\ref{fig:pfm1} implies that JADE assigns all RB resources to the RDD traffic, probably due to its significantly larger packet size (refer to Table~\ref{tbl:traffic}) and the payload-exhaustiveness nature of JADE so that once an RDD UE is selected, it would consume RBs continuously until its payload is met. On the contrary, STAR is able to utilize the resources more efficiently and preserves some RBs for the PD2 UEs. Additionally, the random scheduling fails to make full use of the resources which leads to low TBS.

\noindent 2) As indicated by Figs.~\ref{fig:pfm2} and~\ref{fig:pfm3}, applying exactly the same offline-trained DRL model to different UE traffic ratios, STAR still yields the best TBS performance, which shows its robustness to traffic type distributions. For instance, when only RDD UEs are present, as shown in Fig.~\ref{fig:pfm2}, the resource utilization for both STAR and JADE reaches 100\%, but STAR manages to allocate the RBs more wisely so as to yield higher TBS than JADE. The random scheduling scheme produces the lowest TBS due to inefficient usage of the resources, which is consistent with the observation from Fig.~\ref{fig:pfm1}. On the other hand, when only PD2 UEs exist (see Fig.~\ref{fig:pfm3}), JADE can still yield the largest TBS as compared with the other two algorithms. 

\noindent 3) Combining the simulation results in Figs.~\ref{fig:pfm1}-\ref{fig:pfm3} and the complexity comparison in Table~\ref{tbl:complexity}, it is evident that STAR surpasses JADE in terms of sum-TBS while having significantly lower online computational complexity, thus substantially outperforming JADE considering both system-level performance and online computational complexity. Therefore, it is advantageous to adopt STAR in practice to conduct joint multi-UE and resource scheduling with contiguous FDRA. 

\section{Conclusion}
We have proposed a DRL-based algorithm that jointly schedules UE and contiguous frequency-domain resources. In the proposed method, a DQN agent module is trained offline to determine which UE and how many RBs for that UE should be scheduled at each allocation step. We design the state space, action space, and reward function which are suitable for both the joint UE and resource allocation problem and DRL implementation. System-level simulations demonstrate that compared to a practical contiguous FDRA algorithm named JADE proposed in~\cite{Sun20Fdra} which outperforms prior existing methods, the proposed algorithm enjoys both better performance and significantly lower online computational complexity. This work can be extended to the scenarios which may contain more UEs, more diverse traffic types, and/or other scheduling metrics.

\bibliographystyle{my_spmpsci}  
\bibliography{DRL_Type-1_FDRA}

\end{document}